\documentclass[aps,prl,twocolumn,groupedaddress]{revtex4}
\usepackage[dvips]{graphics, color}

\begin{document}

\title{Observation of resonance condensation of fermionic atom pairs}

\author{C. A. Regal, M. Greiner, and D. S. Jin}
\thanks{Quantum Physics Division, National Institute of Standards and Technology.}
\address{JILA, National Institute of Standards and Technology and
University of Colorado, and Department of Physics, University of
Colorado, Boulder, CO 80309-0440}

\date{January 13, 2004}

\begin{abstract}We have observed condensation of fermionic atom pairs
in  the BCS-BEC crossover regime.  A trapped gas of fermionic
$^{40}$K atoms is evaporatively cooled to quantum degeneracy and
then a magnetic-field Feshbach resonance
 is used to control the atom-atom interactions.  The location of this resonance is
 precisely determined from low-density measurements of molecule dissociation.
 In order to search for condensation on either side of the resonance
 we introduce a technique that pairwise projects fermionic atoms onto molecules;
 this enables us to measure the momentum distribution of fermionic atom pairs.
 The transition to condensation of fermionic atom pairs is mapped out as a function
 of the initial atom gas temperature $T$ compared to the Fermi temperature $T_F$ for
 magnetic-field detunings on both the BCS and BEC sides of the resonance.
 \end{abstract}

\pacs{03.75.Ss, 05.30.Fk}

\maketitle

Ultracold quantum gases of fermionic atoms with tunable
interactions offer the unique possibility to experimentally access
the predicted crossover between BCS-type superfluidity of momentum
pairs and Bose-Einstein condensation (BEC) of molecules
\cite{Leggett1980,Nozieres1985,Randeria1995,Holland2001a,Timmermans2001a,Ohashi2002a,Stajic2003a}.
Magnetic-field Feshbach resonances provide the means for
controlling both the magnitude of cold atom interactions,
characterized by the s-wave scattering length $a$, as well as
whether they are, in the mean-field approximation, effectively
repulsive $(a>0)$ or attractive $(a<0)$
\cite{Stwalley1976b,Tiesinga1993a}. For magnetic-field detunings
on the $a>0$, or BEC, side of the resonance there exists an
extremely weakly bound molecular state whose binding energy
depends strongly on the detuning from the Feshbach resonance
\cite{Donley2002a,Regal2003c}.  In Fermi gases this state can be
long lived
\cite{Strecker2003a,Cubizolles2003a,Jochim2003a,Regal2003d}. BEC
of these molecules represents one extreme of the predicted BCS-BEC
crossover and recently has been observed for both $^{40}$K$_2$ and
$^6$Li$_2$ molecules
\cite{Greiner2003b,Jochim2003b,Zwierlein2003b}.

In atomic Fermi gas systems condensates have not previously been
observed beyond this molecular BEC extreme \cite{note}.  In
discussing condensation of a Fermi gas throughout the BCS-BEC
crossover, terms such as Cooper pairs, molecules, BEC, and
fermionic condensates often have ambiguous meanings.  In this
paper we define ``condensation of fermionic atom pairs", or
equivalently fermionic condensates, as a condensation (i.e. the
macroscopic occupation of a single quantum state) in which the
underlying Fermi statistics of the paired particles plays an
essential role \cite{Stoof1996a}. In the BCS extreme this is more
commonly termed condensation of Cooper pairs.  Fermionic
condensates are distinct from the BEC extreme where there remains
no fermionic degree of freedom because all fermions are bound into
bosonic molecules \cite{moleculenote}. In this Letter we report
the observation of condensation of fermionic atom pairs near and
on both sides of the Feshbach resonance, which corresponds to the
BCS-BEC crossover regime.  We observe condensation on the $a<0$,
or BCS, side of the Feshbach resonance. Here the two-body physics
of the resonance no longer supports the weakly bound molecular
state; hence, only cooperative many-body effects can give rise to
this condensation of fermion pairs
\cite{Holland2001a,Timmermans2001a,Ohashi2002a,Stajic2003a,Duine2003a}.

\begin{figure}[b] \begin{center}
\scalebox{.95}[.95]{\includegraphics{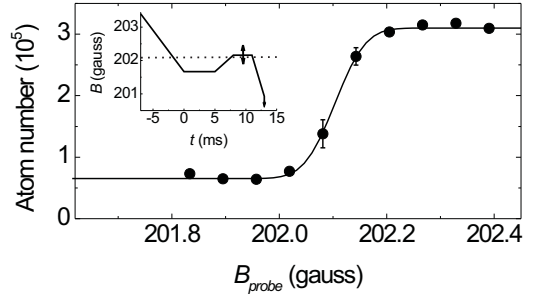}}
\caption{Measurement of the Feshbach resonance position $B_0$.
Shown in the inset is a schematic of the magnetic field as a
function of time $t$ measured with respect to the optical trap
turn off at $t=0$. Molecules are first created by a slow
magnetic-field sweep across the resonance (dotted line) and then
dissociated if $B_{probe}$ (indicated by the arrow in the inset)
is beyond the magnetic field where the two-body physics supports a
new bound state. The number of atoms, measured at $t=17$ ms, is
shown as a function of $B_{probe}$. The two error bars indicate
the spread in repeated points at these values of $B$. A fit of the
data to an error function reveals $B_0=202.10 \pm 0.07$ G, where
the uncertainty is given conservatively by the $10\% - 90\%$
width.} \label{fig1}
\end{center}
\end{figure}

Demonstrating condensation of fermionic atom pairs on the BCS side
of the resonance presents significant challenges.  Observation of
pairing of fermions \cite{Greiner2003a} is insufficient to
demonstrate condensation, and rather a probe of the momentum
distribution is required \cite{Stajic2003a}.  For example, the
standard technique developed for observing BEC relies on
time-of-flight expansion images \cite{Anderson1995a,Davis1995b}.
However, this method is problematic on the BCS side of the
resonance because the pairs depend on many-body effects and are
not bound throughout expansion of the gas.  In this work we
introduce a technique that takes advantage of the Feshbach
resonance to pairwise project the fermionic atoms onto molecules.
We probe the system by rapidly sweeping the magnetic field to the
$a>0$, or BEC, side of the resonance, where time-of-flight imaging
can be used to measure the momentum distribution of the weakly
bound molecules. The projecting magnetic-field sweep is completed
on a timescale that allows molecule formation but is still too
brief for particles to collide or move significantly in the trap.
This projection always results in 60$\%$ to 80$\%$ of the atom
sample appearing as molecules.  However, we find that there is a
threshold curve of $T/T_F$ versus detuning from the Feshbach
resonance below which we observe a fraction of the molecules to
have near zero momentum.  We interpret this as reflecting a
pre-existing condensation of fermionic atom pairs.

Our basic experimental procedures have been discussed in prior
work \cite{DeMarco1999a,Regal2003b}. We trap and cool a dilute gas
of the fermionic isotope $^{40}$K, which has a total atomic spin
$f=9/2$ in its lowest hyperfine ground state and thus ten
available Zeeman spin-states $|f,m_f\rangle$
\cite{DeMarco1999a,Roati2002a}. We use a far-off resonance optical
dipole trap that can confine atoms in any spin state as well as
the molecules we create from these atoms. The optical trap is
characterized by radial frequencies ranging between $\nu_r = 320$
and $440$ Hz, with the trap aspect ratio, $\nu_r/\nu_z$, fixed at
$79 \pm 15$.

\begin{figure} \begin{center}
\scalebox{.85}[.85]{\includegraphics{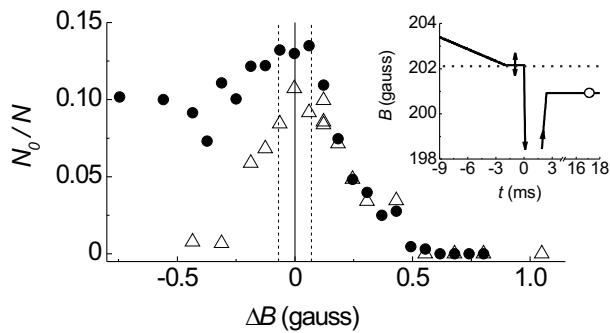}}
\caption{Measured condensate fraction as a function of detuning
from the Feshbach resonance $\Delta B=B_{hold}-B_0$.  Data here
were taken for $t_{hold}=2$ ms ({\large $\bullet$}) and
$t_{hold}=30$ ms ($\triangle$) with an initial cloud at
$T/T_F=0.08$ and $T_F$=0.35 $\mu$K. The area between the dashed
lines around $\Delta B=0$ reflects the uncertainty in the Feshbach
resonance position based on the $10\% - 90\%$ width of the feature
in Fig. \ref{fig1}. Condensation of fermionic atom pairs is seen
near and on either side of the Feshbach resonance. Comparison of
the data taken with the different hold times indicates that the
pair condensed state has a significantly longer lifetime near the
Feshbach resonance and on the BCS ($\Delta B>0$) side.  The inset
shows a schematic of a typical magnetic-field sweep used to
measure the fermionic condensate fraction.  The system is first
prepared by a slow magnetic-field sweep towards the resonance
(dotted line) to a variable position $B_{hold}$, indicated by the
two-sided arrow. After a time $t_{hold}$ the optical trap is
turned off and the magnetic field is quickly lowered by $\sim 10$
G to project the atom gas onto a molecular gas. After free
expansion, the molecules are imaged on the BEC side of the
resonance (${\Large \circ}$). } \label{fig2}
\end{center}
\end{figure}

\begin{figure} \begin{center}
\scalebox{.4}[.4]{\includegraphics{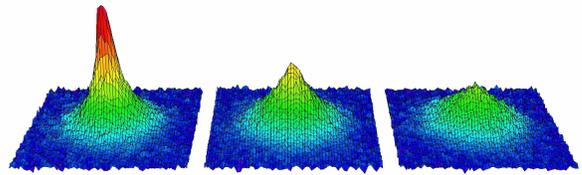}} \caption{Time of
flight images showing the fermionic condensate. The images, taken
after the projection of the fermionic system onto a molecular gas,
are shown for $\Delta B = 0.12,$ 0.25, and 0.55 G (left to right)
on the BCS side of the resonance. The original atom cloud starts
at $T/T_F=0.07$, and the resulting fitted condensate fractions are
$N_0/N = 0.10,$ $0.05$, and $0.01$ (left to right). Each image
corresponds to $N=100,000$ particles and is an average over 10
cycles of the experiment.} \label{fig4}
\end{center}
\end{figure}

Experiments are initiated by preparing atoms in a nearly equal,
incoherent mixture of the $|9/2,-7/2\rangle$ and
$|9/2,-9/2\rangle$ spin states at a low $T/T_F$.  We access an
s-wave Feshbach resonance between these states located at a
magnetic field near 200 G \cite{Bohn2000a}.   A precise
determination of the magnetic-field location of the two-body
resonance is an essential ingredient for exploring the BCS-BEC
crossover regime. In our previous work the location of the
resonance was determined from the peak in the resonantly enhanced
elastic collision rate \cite{Loftus2002a,Regal2003a,Regal2003d}.
In the work reported here we have more precisely determined the
location of the resonance by measuring the magnetic field, $B_0$,
above which the two-body physics no longer supports the shallow
bound state.

Figure \ref{fig1} shows the result of such a measurement.
Molecules created by a slow magnetic-field sweep across the
resonance are dissociated by raising the magnetic field to a value
$B_{probe}$ near the resonance (inset to Fig. 1). To avoid
many-body effects, this dissociation occurs after allowing the gas
to expand from the trap to much lower density. After a total
expansion time of 17 ms atoms not bound in molecules are
selectively detected at near zero field \cite{Regal2003c}. The
measured number of atoms increases sharply at $B_0=202.10 \pm
0.07$ G. This more precise measurement of the resonance position
agrees well with previous results
\cite{Loftus2002a,Regal2003a,Regal2003d}. As an additional check,
we have located the resonance by creating, rather than
dissociating, molecules. The measured number of molecules
decreases sharply at $B=202.14 \pm 0.11$ G in good agreement with
the molecule dissociation result \cite{rfnote}.

In order to investigate the BCS-BEC crossover regime we initially
prepare the ultracold two-component atom gas at a magnetic field
of 235.6 G, far above the Feshbach resonance.  Here the gas is not
strongly interacting, and we measure $T/T_F$ through surface fits
to time-of-flight images of the Fermi gas
\cite{DeMarco1999a,Regal2003b}. The field is then slowly lowered
at typically 10 ms/G to a value $B_{hold}$ near the resonance.
This sweep is slow enough to allow the atoms and molecules
sufficient time to move and collide in the trap.  This was shown
previously in Ref. \cite{Greiner2003b} where, for a Fermi gas
initially below $T/T_F=0.17$, a magnetic-field sweep at 10 ms/G to
a final $B$ 0.56 G below the resonance produced a molecular
condensate. In this Letter, we now explore the behavior of the
sample when sweeping slowly to values of $B_{hold}$ on either side
of the Feshbach resonance.

To probe the system we pairwise project the fermionic atoms onto
molecules and measure the momentum distribution of the resulting
molecular gas.  This projection is accomplished by rapidly
lowering the magnetic field by $\sim 10$ G at a rate of typically
$(50$ $\mu {\rm s/G})^{-1}$ while simultaneously releasing the gas
from the trap. This puts the gas far on the BEC side of the
resonance, where it is weakly interacting.   The total number of
molecules after the projection, $N$, corresponds to 60$ \%$ to
80$\%$ of the original atom number in each spin state. After a
total of typically 17 ms of expansion the molecules are
selectively detected using rf photodissociation immediately
followed by spin-selective absorption imaging \cite{Greiner2003b}.
To look for condensation, these absorption images are surface fit
to a two-component function that is the sum of a Thomas-Fermi
profile for a condensate and a gaussian function for non-condensed
molecules \cite{Greiner2003b}.

\begin{figure} \begin{center}
\scalebox{.4}[.4]{\includegraphics{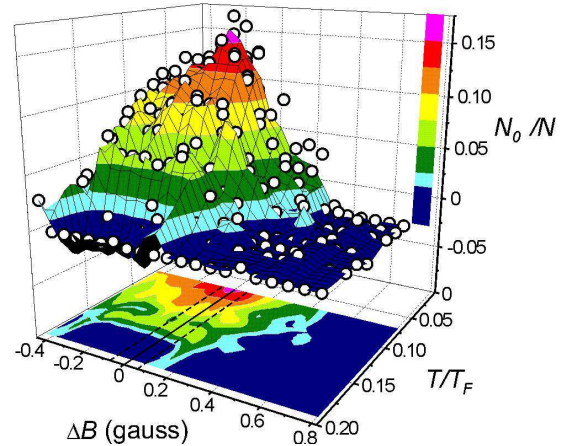}}
\caption{Transition to condensation as a function of both $\Delta
B$ and $T/T_F$. The data for this phase diagram were collected
with the same procedure as shown in the inset to Fig. 2 with
$t_{hold} \sim 2$ ms.  The area between the dashed lines around
$\Delta B=0$ reflects the uncertainty in the Feshbach resonance
location from the width of the feature in Fig. 1.  The false color
surface and contour plot are obtained using a Renka-Cline
interpolation of approximately 200 distinct data points (${\Large
\circ}$) \cite{Renka1984}. One measure of when the gas becomes
strongly interacting is the criterion $|k_F a|
> 1$, where $\hbar k_F$ is the Fermi momentum
\cite{O'Hara2002a,Regal2003b,Bourdel2003a,Gupta2003b}. For these
data, $|\Delta B| < 0.6$ corresponds to $|k_F a|
> 1$.} \label{fig3}
\end{center}
\end{figure}

Figures \ref{fig2} through \ref{fig3} present the main result of
this paper. In Fig. \ref{fig2} we plot the measured condensate
fraction $N_0/N$ as a function of the magnetic-field detuning from
the resonance, $\Delta B=B_{hold}-B_0$ \cite{width}.  The data in
Fig. \ref{fig2} was taken for a Fermi gas initially at
$T/T_F=0.08$ and for two different wait times at $B_{hold}$.
Condensation is observed on both the BCS ($\Delta B>0$) and BEC
($\Delta B<0$) sides of the resonance. We further find that the
condensate on the BCS side of the Feshbach resonance has a
relatively long lifetime.  The lifetime was probed by increasing
$t_{hold}$ to 30 ms (triangles in Fig. \ref{fig2}). We find that
for the BEC side of the resonance no condensate is observed for
$t_{hold}=30$ ms except very near the resonance. However, for all
data on the BCS side of the resonance the observed condensate
fraction is still $>70\%$ of that measured for $t_{hold}=2$ ms.
Finally, we note that the appearance of the condensate is
accompanied by a significant (as large as $20\%$ at the resonance)
decrease in the measured width of the non-condensed fraction. This
effect will be a subject of future investigations.

Figure \ref{fig4} displays sample time-of-flight absorption images
for the fermionic condensate.  Figure \ref{fig3} is a phase
diagram created from our data; here we plot the measured
condensate fraction as function of $\Delta B$ and of the initial
Fermi gas degeneracy $T/T_F$. The condensate forms at lower
initial $T/T_F$ with increasing $\Delta B$, an effect predicted in
\cite{Holland2001a,Timmermans2001a,Ohashi2002a,Stajic2003a}.

An essential aspect of these measurements is the fast
magnetic-field sweep that pairwise projects the fermionic atoms
onto molecules. It is a potential concern that the condensation
might occur during this sweep rather than at $B_{hold}$. However,
in our previous work it was shown that a magnetic-field sweep with
an inverse speed less than $800$ $\mu {\rm s/G}$ was too fast to
produce a molecular condensate when starting with a Fermi gas 0.68
G on the BCS side of the resonance \cite{Greiner2003b}. Thus, the
inverse sweep speed we use in this Letter of typically $50$ $\mu
{\rm s/G}$, while sufficiently slow to convert $60\%$ of the
sample to weakly bound molecules \cite{Regal2003c}, is much too
fast to produce a molecular condensate.

\begin{figure} \begin{center}
\scalebox{1.6}[1.6]{\includegraphics{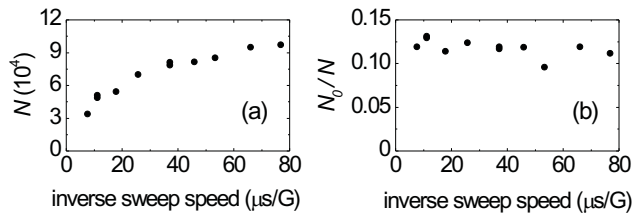}}
\caption{Dependence of molecule number and condensate fraction on
the speed of the fast magnetic-field sweep from the atomic gas
onto the molecular gas. Here $\Delta B=0.12$ and the initial
$T/T_F$ is 0.08. (a) Total number of molecules as a function of
inverse sweep speed. For the fastest sweep speeds fewer molecules
are created, consistent with studies in Ref. \cite{Regal2003c}.
(b) Condensate fraction as a function of the inverse sweep speed.
Even for the fastest sweeps and lowest molecule number, we observe
an unchanged condensate fraction.} \label{fig5}
\end{center}
\end{figure}

In addition, we have checked that the observation of a condensate
on the BCS side of the resonance does not depend on this sweep
speed. As seen in Fig. \ref{fig5}(a), much faster sweeps result in
fewer molecules. This is consistent with our previous study of the
molecule creation process \cite{Regal2003c}. However, we find that
the measured condensate fraction is independent of the sweep rate
(Fig. \ref{fig5}(b)). Even with the lower number of molecules, and
therefore a lower phase space density of the molecular gas, we
observe an essentially unchanged condensate fraction.

Finally we note that, as in our previous measurements performed in
the BEC limit, the measured condensate fraction always remains
well below one \cite{Greiner2003b}.  As part of our probing
procedure the magnetic field is dwells at $\Delta B \sim 10$ G
where the molecule lifetime is only on the order of milliseconds
\cite{Regal2003d,Petrov2003}. This results in a measured loss of
50$\%$ of the molecules and may also reduce the measured
condensate fraction.

In conclusion, we have introduced a method for probing the
momentum distribution of fermionic atom pairs and employed this
technique to observe fermionic condensates near a Feshbach
resonance. By projecting the system onto a molecular gas, we map
out condensation of fermionic pairs as a function of both the
magnetic-field detuning from the resonance and the initial Fermi
degeneracy $T/T_F$.  The fermionic condensates seen in this work
occur in the BCS-BEC crossover regime, far from the perturbative
BCS limit. As predicted, the system is observed to vary smoothly
in the BCS-BEC crossover regime. Further, the lifetime of the
condensed state is found to be significantly longer in the
crossover regime than it is in the BEC limit. As in the case of
BEC, one expects the resonance fermionic condensation observed
here to correspond to superfluidity.  The experimental realization
of condensation in the BCS-BEC crossover regime demonstrated in
this Letter follows more than two decades of theoretical
investigation and initiates experimental study of this physics.

We thank E. A. Cornell, C. E. Wieman, M. Holland, K. Levin, E.
Altman, and L. Radzihovsky for stimulating discussion and J. T.
Smith for experimental assistance. This work was supported by NSF
and NIST; C. A. R. acknowledges support from the Hertz Foundation.

%\bibliographystyle{prsty}
%\bibliography{revised}

\begin{thebibliography}{10}

\bibitem{Leggett1980}
A.~J. Leggett, J. Phys. C. (Paris) {\bf 41},  7  (1980).

\bibitem{Nozieres1985}
P. Nozieres and S. Schmitt-Rink, J. Low-Temp. Phys. {\bf 59},  195
(1985).

\bibitem{Randeria1995}
M. Randeria,  in {\em Bose-Einstein Condensation}, edited by A.
Griffin, D.~W.
  Snoke, and S. Stringari (Cambridge University, Cambridge, 1995), pp.\
  355--392.

\bibitem{Holland2001a}
M. Holland, S.~J.~J.~M.~F. Kokkelmans, M.~L. Chiofalo, and R.
Walser, Phys.
  Rev. Lett. {\bf 87},  120406  (2001).

\bibitem{Timmermans2001a}
E. Timmermans, K. Furuya, P.~W. Milonni, and A.~K. Kerman, Phys.
Lett. A {\bf
  285},  228  (2001).

\bibitem{Ohashi2002a}
Y. Ohashi and A. Griffin, Phys. Rev. Lett. {\bf 89},  130402
(2002).

\bibitem{Stajic2003a}
J. Stajic {\it et~al.}, cond-mat/0309329  (2003).

\bibitem{Stwalley1976b}
W.~C. Stwalley, Phys. Rev. Lett. {\bf 37},  1628  (1976).

\bibitem{Tiesinga1993a}
E. Tiesinga, B.~J. Verhaar, and H.~T.~C. Stoof, Phys. Rev. A {\bf
47},  4114
  (1993).

\bibitem{Donley2002a}
E.~A. Donley, N.~R. Claussen, S.~T. Thompson, and C.~E. Wieman,
Nature {\bf
  417},  529  (2002).

\bibitem{Regal2003c}
C.~A. Regal, C. Ticknor, J.~L. Bohn, and D.~S. Jin, Nature {\bf
424},  47
  (2003).

\bibitem{Strecker2003a}
K.~E. Strecker, G.~B. Partridge, and R.~G. Hulet, Phys. Rev. Lett.
{\bf 91},
  080406  (2003).

\bibitem{Cubizolles2003a}
J. Cubizolles {\it et~al.}, Phys. Rev. Lett. {\bf 91},  240401
(2003).

\bibitem{Jochim2003a}
S. Jochim {\it et~al.}, Phys. Rev. Lett {\bf 91},  240402  (2003).

\bibitem{Regal2003d}
C.~A. Regal, M. Greiner, and D.~S. Jin, Phys. Rev. Lett. in press
(2004).

\bibitem{Greiner2003b}
M. Greiner, C.~A. Regal, and D.~S. Jin, Nature {\bf 426},  537
(2003).

\bibitem{Jochim2003b}
S. Jochim {\it et~al.}, Science {\bf 302},  2101  (2003).

\bibitem{Zwierlein2003b}
M.~W. Zwierlein {\it et~al.}, Phys. Rev. Lett. {\bf 91},  250401
(2003).

\bibitem{note}
Several groups have reported attaining Fermi gases on the BCS side
of a
  Feshbach resonance at $T/T_F$ below theoretical predictions for the critical
  condensation temperature; so far there has been no experimental evidence for
  condensate formation
  \cite{O'Hara2002a,Regal2003b,Strecker2003a,Cubizolles2003a,Bartenstein2004a}.

\bibitem{Stoof1996a}
H.~T.~C. Stoof, M. Houbiers, C.~A. Sackett, and R.~G. Hulet, Phys.
Rev. Lett.
  {\bf 76},  10  (1996).

\bibitem{moleculenote}
We reserve the term molecule for a two-body bound state.

\bibitem{Duine2003a}
See also:  R.~A. Duine and H.~T.~C. Stoof, J. Opt. B {\bf 5},
S212 (2003).

\bibitem{Greiner2003a}
M. Greiner {\it et~al.}, cond-mat/0308519  (2003).

\bibitem{Anderson1995a}
M.~H. Anderson {\it et~al.}, Science {\bf 269},  198  (1995).

\bibitem{Davis1995b}
K.~B. Davis {\it et~al.}, Phys. Rev. Lett. {\bf 75},  3969
(1995).

\bibitem{DeMarco1999a}
B. DeMarco and D.~S. Jin, Science {\bf 285},  1703  (1999).

\bibitem{Regal2003b}
C.~A. Regal and D.~S. Jin, Phys. Rev. Lett. {\bf 90},  230404
(2003).

\bibitem{Roati2002a}
G. Roati, F. Riboli, G. Modugno, and M. Inguscio, Phys. Rev. Lett.
{\bf 89},
  150403  (2002).

\bibitem{Bohn2000a}
J.~L. Bohn, Phys. Rev. A {\bf 61},  053409  (2000).

\bibitem{Loftus2002a}
T. Loftus {\it et~al.}, Phys. Rev. Lett. {\bf 88},  173201
(2002).

\bibitem{Regal2003a}
C.~A. Regal, C. Ticknor, J.~L. Bohn, and D.~S. Jin, Phys. Rev.
Lett. {\bf 90},
  053201  (2003).

\bibitem{rfnote}
We calibrate the magnetic field for gases in the trap and during
expansion
  using rf transitions between Zeeman levels. The magnetic field is
  reproducible to $<15$ mG. Mean-field shifts near the resonance are avoided
  through transfer between the resonant states \cite{Zwierlein2003a}.

\bibitem{width}
The scattering length corresponding to $\Delta B$ can be
calculated from
  $a=a_{bg}(1-\frac{w}{\Delta B})$, where $a_{bg}=174$ $a_0$ and $w=7.8 \pm
  0.6$ G \cite{Regal2003a}.

\bibitem{Renka1984}
R.~L. Renka and A.~K. Cline, Rocky Mountain J. Math {\bf 14},  223
(1984).

\bibitem{O'Hara2002a}
K.~M. O'Hara {\it et~al.}, Science {\bf 298},  2179  (2002).

\bibitem{Bourdel2003a}
T. Bourdel {\it et~al.}, Phys. Rev. Lett. {\bf 91},  020402
(2003).

\bibitem{Gupta2003b}
S. Gupta {\it et~al.}, Science {\bf 300},  1723  (2003).

\bibitem{Petrov2003}
D.~S. Petrov, C. Salomon, and G.~V. Shlyapnikov, cond-mat/0309010
(2003).

\bibitem{Bartenstein2004a}
M. Bartenstein {\it et~al.}, cond-mat/0401109  (2004).

\bibitem{Zwierlein2003a}
M.~W. Zwierlein, Z. Hadzibabic, S. Gupta, and W. Ketterle, Phys.
Rev. Lett.
  {\bf 91},  250404  (2003).

\end{thebibliography}

\end{document}